\documentclass[12pt,english]{article}
\usepackage[T1]{fontenc}
\usepackage[latin1]{inputenc}
\usepackage{geometry}
\usepackage{amssymb,amsmath}

\usepackage{babel}
\usepackage{graphics}

\renewcommand{\thefootnote}{\fnsymbol{footnote}}
\newlength{\pubnumber} 

\catcode`\@=11
\@addtoreset{equation}{section}

\def\section{\@startsection{section}{1}{\z@}{3.5ex plus 1ex minus .2ex}
 {2.3ex plus .2ex}{\large\bf}}
\def\subsection{\@startsection{subsection}{2}{\z@}{2.3ex plus .2ex}
 {2.3ex plus .2ex}{\bf}}

\usepackage{amssymb}
\usepackage{epsfig}
\usepackage{cite}
\newcommand{\ba}{\begin{eqnarray}}
\newcommand{\ea}{\end{eqnarray}}

\begin{document}
\begin{titlepage}
\samepage{
\setcounter{page}{1}
\rightline{LTH--856}
\vfill
\begin{center}
 {\Large \bf  Interpolations Among \\
           NAHE--based Supersymmetric \\
        and Nonsupersymmetric String Vacua\\}
\vfill
 {\large  Alon E. Faraggi \footnote{
	E-mail address: faraggi@amtp.liv.ac.uk}
and
	  Mirian Tsulaia $\setcounter{footnote}{4}$\footnote{
        E-mail address: tsulaia@liv.ac.uk}
\\
\vspace{.4in}
}

 {\it           Department of Mathematical Sciences,
		University of Liverpool,     \\
                Liverpool L69 7ZL, United Kingdom}\\
\end{center}
\vfill
\begin{abstract}
  {\rm 
The quasi--realistic free fermionic heterotic--string models provide some of 
the most detailed examples to explore the phenomenology of string unification.
While providing a powerful tool to generate models and their spectra, 
understanding the realisation of the free fermion models in a bosonic
formalism will provide important insight into their basic properties away from
the free fermion point. In this paper we elucidate bosonic equivalent
of the basic 
symmetry breaking pattern in the free fermion models from $E_8\times E_8$ to
$SO(16)\times SO(16)$ and  exhibit the connection of the 
free fermion models with corresponding non--supersymmetric 
vacua  by interpolations.
}
\end{abstract}
\smallskip}
\end{titlepage}

\renewcommand{\thefootnote}{\arabic{footnote}}
\setcounter{footnote}{0}

\def\l{\label}
\def\beq{\begin{equation}}
\def\eeq{\end{equation}}
\def\beqn{\begin{eqnarray}}
\def\eeqn{\end{eqnarray}}

\def\ie{{\it i.e.}}
\def\eg{{\it e.g.}}
\def\half{{\textstyle{1\over 2}}}
\def\third{{\textstyle {1\over3}}}
\def\quarter{{\textstyle {1\over4}}}
\def\m{{\tt -}}
\def\p{{\tt +}}

\def\slash#1{#1\hskip-6pt/\hskip6pt}
\def\slk{\slash{k}}
\def\GeV{\,{\rm GeV}}
\def\TeV{\,{\rm TeV}}
\def\y{\,{\rm y}}
\def\SM{Standard-Model }
\def\SUSY{supersymmetry }
\def\SSSM{supersymmetric standard model}
\def\vev#1{\left\langle #1\right\rangle}
\def\l{\langle}
\def\r{\rangle}

\def\Htw{{\tilde H}}
\def\chibar{{\overline{\chi}}}
\def\qbar{{\overline{q}}}
\def\ibar{{\overline{\imath}}}
\def\jbar{{\overline{\jmath}}}
\def\Hbar{{\overline{H}}}
\def\Qbar{{\overline{Q}}}
\def\abar{{\overline{a}}}
\def\alphabar{{\overline{\alpha}}}
\def\betabar{{\overline{\beta}}}
\def\tautwo{{ \tau_2 }}
\def\thetatwo{{ \vartheta_2 }}
\def\thetathree{{ \vartheta_3 }}
\def\thetafour{{ \vartheta_4 }}
\def\ttwo{{\vartheta_2}}
\def\tthree{{\vartheta_3}}
\def\tfour{{\vartheta_4}}
\def\ti{{\vartheta_i}}
\def\tj{{\vartheta_j}}
\def\tk{{\vartheta_k}}
\def\calF{{\cal F}}
\def\smallmatrix#1#2#3#4{{ {{#1}~{#2}\choose{#3}~{#4}} }}
\def\ab{{\alpha\beta}}
\def\Minv{{ (M^{-1}_\ab)_{ij} }}
\def\bone{{\bf 1}}
\def\ii{{(i)}}
\def\V{{\bf V}}
\def\b{{\bf b}}
\def\N{{\bf N}}
\def\t#1#2{{ \Theta\left\lbrack \matrix{ {#1}\cr {#2}\cr }\right\rbrack }}
\def\C#1#2{{ C\left\lbrack \matrix{ {#1}\cr {#2}\cr }\right\rbrack }}
\def\tp#1#2{{ \Theta'\left\lbrack \matrix{ {#1}\cr {#2}\cr }\right\rbrack }}
\def\tpp#1#2{{ \Theta''\left\lbrack \matrix{ {#1}\cr {#2}\cr }\right\rbrack }}
\def\l{\langle}
\def\r{\rangle}

\def\La{\Lambda}
\def\te{\theta}


\def\inbar{\,\vrule height1.5ex width.4pt depth0pt}

\def\IC{\relax\hbox{$\inbar\kern-.3em{\rm C}$}}
\def\IQ{\relax\hbox{$\inbar\kern-.3em{\rm Q}$}}
\def\IR{\relax{\rm I\kern-.18em R}}
 \font\cmss=cmss10 \font\cmsss=cmss10 at 7pt
\def\IZ{\relax\ifmmode\mathchoice
 {\hbox{\cmss Z\kern-.4em Z}}{\hbox{\cmss Z\kern-.4em Z}}
 {\lower.9pt\hbox{\cmsss Z\kern-.4em Z}}
 {\lower1.2pt\hbox{\cmsss Z\kern-.4em Z}}\else{\cmss Z\kern-.4em Z}\fi}

\def\AEF{A.E. Faraggi}
\def\NPB#1#2#3{{Nucl.\ Phys.}\/ {B \bf #1} (#2) #3}
\def\PLB#1#2#3{{Phys.\ Lett.}\/ {B \bf #1} (#2) #3}
\def\PRD#1#2#3{{Phys.\ Rev.}\/ {D \bf #1} (#2) #3}
\def\PRL#1#2#3{{Phys.\ Rev.\ Lett.}\/ {\bf #1} (#2) #3}
\def\PRP#1#2#3{{Phys.\ Rep.}\/ {\bf#1} (#2) #3}
\def\MODA#1#2#3{{Mod.\ Phys.\ Lett.}\/ {\bf A#1} (#2) #3}
\def\IJMP#1#2#3{{Int.\ J.\ Mod.\ Phys.}\/ {A \bf #1} (#2) #3}
\def\nuvc#1#2#3{{Nuovo Cimento}\/ {\bf #1A} (#2) #3}
\def\JHEP#1#2#3{{JHEP} {\textbf #1}, (#2) #3}
\def\EJP#1#2#3{{\it Eur.\ Phys.\ Jour.}\/ {\bf C#1} (#2) #3}
\def\MPLA#1#2#3{{\it Mod.\ Phys.\ Lett.}\/ {\bf A#1} (#2) #3}
\def\IJMPA#1#2#3{{\it Int.\ J.\ Mod.\ Phys.}\/ {\bf A#1} (#2) #3}

\def\etal{{\it et al\/}}

\hyphenation{su-per-sym-met-ric non-su-per-sym-met-ric}
\hyphenation{space-time-super-sym-met-ric}
\hyphenation{mod-u-lar mod-u-lar--in-var-i-ant}


\setcounter{footnote}{0}
\section{Introduction}
\bigskip

The heterotic--string models in the free fermionic formulation
are among the most realistic string models constructed to date \cite{review}. 
Indeed, the quasi--realistic models utilizing this formalism,
constructed nearly two decades ago \cite{fsu5,slm,alr},
have been utilised to explore 
how many of the issues pertaining to the phenomenology of the
Standard Model and Grand Unified
Theories may arise from string theory, including: 
fermion masses, mixing and CP--violation;
neutrino masses; 
proton stability;
gauge coupling unification;
supersymmetry breaking and smatter degeneracy \cite{review}.
Additionally, this construction
gives rise to models in which the Standard Model charged spectrum 
below the string scale consist solely of that of the Minimal Supersymmetric
Standard Model \cite{slm}.
The phenomenological free fermionic string models therefore serve as
a laboratory in which we can investigate how the 
Standard Model data may be obtained from a fundamental theory. 
In turn, these models are used to reveal general
properties of string theory. In this vein, duality under the 
exchange of spinors and vectors of the $SO(10)$ GUT group was 
discovered \cite{fkr}. This, and other, duality symmetries in the string vacua 
space, indicate that the string models live in a connected space.
Thus, while from point of view of the effective low energy field
theory limit of the string vacua, the spinor--vector dual pairs
correspond to nonequivalent theories, the vacua are connected
in string theory. Essentially, we can say that in string theory
the duality map interchanges between massless and massive string 
states, hence inducing the duality map between vacua that in the 
effective low energy field theory correspond to completely different
phenomenology. 

The need to understand better the properties of the free fermionic models
is evident. An important 
feature of the realistic free fermionic models is their underlying
$Z_2\times Z_2$ orbifold structure \cite{foc,partitions}.
Furthermore, a vital  distinction of the
quasi--realistic free fermionic models 
is that the initial gauge symmetry at the level of the
toroidal compactification with 
four dimensional $N=4$ space--time supersymmetry, 
which is then moded by the $Z_2\times Z_2$ orbidold to $N=1$, 
correspond to $SO(16)\times SO(16)$, rather than the more conventional
$E_8\times E_8$ symmetry, which is usually the initial symmetry 
in orbifold constructions. This distinction has important 
phenomenological consequence that have been elaborated 
elsewhere \cite{higgsmattersplit}. In the free fermionic formalism the two cases 
can be seen to arise from a discreet choice of a Generalised 
GSO (GGSO) projection coefficient. Alternatively,
as has been demonstrated in \cite{partitions} by studying
the respective partition functions,
the two cases can be connected by a $Z_2\times Z_2$ orbifold.
In the ten dimensional
case the relevant GGSO phase correspond to the 
discreet choice between the $N=1$ supersymmetric heterotic 
$E_8\times E_8$ string \cite{heterotic} versus the non--supersymmetric 
$SO(16)\times SO(16)$ heterotic--string vacuum \cite{so16so16}.

In the free fermionic formalism the relevant phase is the 
GGSO phase, which is responsible for the projection
of the spinorial representations in the decomposition of
the adjoint of $E_8\times E_8$ under $SO(16)\times SO(16)$ 
\cite{foc,partitions},
hence inducing the reduction of the symmetry from  
$E_8\times E_8$ to  $SO(16)\times SO(16)$. 
The discreet GGSO choice is responsible for this projection in the
ten and four dimensional free fermion models. Alternatively, one can
formulate the same projection at the free fermion point
as orbifolding by a fermion number that acts on the gauge
degrees of freedom coupled with a fermion number acting on the internal
lattice, in the four dimensional case \cite{partitions}, or on the space--time 
degrees of freedom in the ten dimensional case \cite{so16so16}. 
The natural question is what is the relation between the cases. 
In this paper we address this question by writing explicitly the partition 
function of these vacua at the free fermion point,
and subsequently at the general point of moduli space.
This may assist on better understanding of the internal
spaces of these vacua and possibly the geometrical
structure that they give rise to.

We first formulate the
orbifold projection from the $E_8\times E_8$ to $SO(16)\times SO(16)$
gauge symmetry as a fermion number acting on the gauge 
degrees of freedom coupled with a shift in the internal
the internal lattice. We then construct the partition functions
corresponding to the model with the internal shift and demonstrate
how it can be interpolated with the 
corresponding nonsupersymmetric vacuum.
The results presented here therefore elucidate the role of the free phase
in the free fermionic models and how its operation is translated to the orbifold 
construction. One can then hope to be able to retrieve some of the successful
phenomenological and duality features of the free fermion models in the
orbifold models. In the reverse direction, one may hope to gain some insight into 
the geometrical structures that underlie
the free fermion models, in particular in 
relation to the dualities and vacuum selection.

Our paper is organised as follows. In section \ref{ffmreview} we
review the structure of the 
quasi--realistic free fermionic models, and in particular the
characteristics that are relevant
for the analysis here. In section \ref{chalukot} we discuss the 
partition functions underlying the NAHE--based free fermionic models. 
In section \ref{interpolazia} we discuss the interpolations among the
NAHE--based partition functions. 
Section \ref{maskanot} concludes the paper.

\section{Review of free fermionic models}\label{ffmreview}

In this section we discuss some
of the relevant features of the free fermionic models
(see \cite{review} for a more detailed introduction).
In the free fermionic formulation of the heterotic string
in four dimensions all the world-sheet
degrees of freedom  required to cancel
the conformal anomaly are represented in terms of free fermions
propagating on the string world-sheet \cite{fff}.
In the light-cone gauge the world-sheet field content consists
of two transverse left- and right-moving space-time coordinate bosons,
$X_{1,2}^\mu$ and ${\bar X}_{1,2}^\mu$,
and their left-moving fermionic superpartners $\psi^\mu_{1,2}$,
and additional 62 purely internal
Majorana-Weyl fermions, of which 18 are left-moving,
$$ \chi^{1,..,6}, \quad y^{1,...,6}, \quad \omega^{1,...,6}$$
and 44 are right-moving 
$${\overline y}^{1,...,6}, \quad {\overline \omega}^{1,...,6}, 
\quad {\overline \psi}^{1,..,5}, \quad {\overline \eta}^{1,2,3}, 
\quad {\overline \phi}^{1,...,8}.$$
Under parallel transport around a non-contractible loop on the toroidal
world-sheet the fermionic fields pick up a phase,
$
f~\rightarrow~-{\rm e}^{i\pi\alpha(f)}f~,~~\alpha(f)\in(-1,+1].
$
Each set of specified
phases for all world-sheet fermions, around all the non-contractible
loops is called the spin structure of the model. Such spin structures
are usually given in the form of 64 dimensional boundary condition vectors,
with each element of the vector specifying the phase of the corresponding
world-sheet fermion. The basis vectors are constrained by string consistency
requirements and completely determine the vacuum structure of the model.
The physical spectrum is obtained by applying the generalised GSO projections.
The boundary condition basis defining a typical 
``realistic free fermionic heterotic string model'' is 
constructed in two stages.
The first stage consists of the NAHE set,
which is a set of five boundary condition basis vectors, 
$\{ 1 ,S,b_1,b_2,b_3\}$ \cite{nahe}
$$ S=\{\psi^{1,2}, \chi^{1,...,6}\}, \quad b_1= \{ \psi^{1,2}, \chi^{1,2}, y^{3,..,6}| {\overline y}^{3,..6}, {\overline  \psi}^{1,..,5,}, {\overline \eta}^{1} \}$$  
	                $$ b_2= \{ \psi^{1,2}, \chi^{3,4}, y^{1,2}, \omega^{5,6}| {\overline y}^{1,2}, {\overline \omega}^{5,6},{\overline  \psi}^{1,..,5,}, {\overline \eta}^{2} \}$$
                        $$  b_3= \{ \psi^{1,2}, \chi^{3,4},  \omega^{1,..,4}|  {\overline \omega}^{1,..,4},{\overline  \psi}^{1,..,5,}, {\overline \eta}^{3} \} $$
where fields with  $\alpha (f)=1$ are indicated explicitly 
The gauge group after imposing the GSO projections induced
by the NAHE set is ${\rm SO} (10)\times {\rm SO}(6)^3\times {\rm E}_8$
with ${N}=1$ supersymmetry.

The second stage of the
construction consists of adding to the 
NAHE set three (or four) additional boundary condition basis vectors,
typically denoted by $\{\alpha,\beta,\gamma\}$. 
These additional basis vectors reduce the number of generations
to three chiral generations, one from each of the sectors $b_1$,
$b_2$ and $b_3$, and simultaneously break the four dimensional
gauge group. 
The assignment of boundary conditions to
$\{{\bar\psi}^{1,\cdots,5}\}$ breaks SO(10) to one of its subgroups
${\rm SU}(5)\times {\rm U}(1)$ \cite{fsu5}, ${\rm SO}(6)\times {\rm SO}(4)$ 
\cite{alr},
${\rm SU}(3)\times {\rm SU}(2)\times {\rm U}(1)^2$ \cite{slm}
or ${\rm SU}(3)\times {\rm SU}(2)^2\times {\rm U}(1)$ \cite{cfs}.

The correspondence of the NAHE-based free fermionic models
with the orbifold construction is illustrated
by extending the NAHE set, $\{ 1,S,b_1,b_2,b_3\}$, by one additional
boundary condition basis vector \cite{foc},
\begin{equation} \label{vectorx}
\xi_1= \{ {\overline \psi}^{1,..,5}, {\overline \eta}^{1,2,3} \},  
\end{equation}
To construct the model in the orbifold formulation one starts
with the compactification on a torus with nontrivial background
fields \cite{Narain}.
The subset of basis vectors
\beq
\{ 1,S,\xi_1,\xi_2\}, \quad \xi_2 = 1+ b_1+b_2 +b_3
\label{neq4set}
\eeq
generates a toroidally-compactified model with ${N}=4$ space-time
supersymmetry and ${\rm SO}(12)\times {\rm E}_8\times {\rm E}_8$ gauge group.
Adding the two basis vectors $b_1$ and $b_2$ to the set
(\ref{neq4set}) corresponds to the ${Z}_2\times {Z}_2$
orbifold model with standard embedding.
Starting from the Narain model with ${\rm SO}(12)\times 
{\rm E}_8\times {\rm E}_8$
symmetry~\cite{Narain}, and applying the ${Z}_2\times {Z}_2$ 
twist on the
internal coordinates, reproduces
the spectrum of the free-fermion model
with the six-dimensional basis set
$\{ 1,S,\xi_1,\xi_2,b_1,b_2\}$.
The Euler characteristic of this model is 48 with $h_{11}=27$ and
$h_{21}=3$.

The effect of the additional basis vector $\xi_1$ of eq.
(\ref{vectorx}), is to separate the gauge degrees of freedom, spanned by
the world-sheet fermions $\{{\bar\psi}^{1,\cdots,5},
{\bar\eta}^{1},{\bar\eta}^{2},{\bar\eta}^{3},{\bar\phi}^{1,\cdots,8}\}$,
from the internal compactified degrees of freedom $\{y,\omega\vert
{\bar y},{\bar\omega}\}^{1,\cdots,6}$. 
In the ``realistic free fermionic
models'' this is achieved by the vector $2\gamma$ \cite{foc}
\begin{equation} \label{vector2gamma}
2\gamma = \{ {\overline \psi}^{1,...,5}, {\overline \eta}^{1,2,3},
{\overline \phi}^{1...,4} \}
\end{equation}
This vector breaks $E_8 \times E_8$ gauge symmetry down to
$SO(16) \times SO(16)$, whereas $Z_2 \times Z_2$ orbifold further
breaks it down to $SO(4)^3 \times SO(10) \times U(1)^3 \times SO(16)$.
The orbifold still yields a model with 24 generations,
eight from each twisted sector,
but now the generations are in the chiral 16 representation
of SO(10), rather than in the 27 of ${\rm E}_6$. The same model can
be realized with the set
$\{ 1,S,\xi_1,\xi_2,b_1,b_2\}$,
by projecting out the $16\oplus{\overline{16}}$
from the $\xi_1$-sector taking
\beq
c{\xi_1\choose \xi_2}\rightarrow -c{\xi_1\choose \xi_2},
\label{changec}
\eeq
where $c{\xi_1\choose \xi_2}$ is a GGSO phase appearing in the partition function.
This choice also projects out the massless vector bosons in the
128 of SO(16) in the hidden-sector ${\rm E}_8$ gauge group, thereby
breaking the ${\rm E}_6\times {\rm E}_8$ symmetry to
${\rm SO}(10)\times {\rm U}(1)\times {\rm SO}(16)$.

The freedom in ({\ref{changec}) corresponds to 
a discrete torsion in the free fermionic model. 
At the level of the ${N}=4$ 
Narain model generated by the set (\ref{neq4set}),
we can define two models, ${Z}_+$ and ${Z}_-$, depending on the sign
of the discrete torsion in eq. (\ref{changec}). The first, say ${Z}_+$,
produces the ${\rm E}_8\times {\rm E}_8$ model, whereas the second, say 
${Z}_-$, produces the ${\rm SO}(16)\times {\rm SO}(16)$ model. 
However, the ${Z}_2\times {Z}_2$
twist acts identically in the two models, and their physical characteristics
differ only due to the discrete torsion eq. (\ref{changec}). 

The projection induced by eqs. (\ref{vector2gamma}), or (\ref{changec}), 
has important phenomenological consequences in the free fermionic 
constructions that are relevant for orbfiold models. In the case of $Z_+$
the action of the $Z_2\times Z_2$ orbifold is to break the observable 
$E_8$ symmetry to $E_6\times U(1)^2$. The chiral matter states are 
contained in the $27$ representation of $E_6$, which decomposes as 
\beq
27= 16_{1\over2}+10_{-1}+1_2\label{27decom}
\eeq
under the $SO(10)\times U(1)$ subgroup of $E_6$, where the spinorial $16$
and vectorial $10$  representations of $SO(10)$ 
contain the Standard Model fermion 
and Higgs states, respectively. The projection induced by (\ref{changec})
in $Z_-$ entails that either the spinorial,  or the vectorial, representation 
survives the GSO projection at a given fixed point. Hence, this projection
operates a Higgs--matter splitting mechanism \cite{higgsmattersplit}
in the phenomenological free fermionic models. 

In contra--distinction 
most of the heterotic orbifold models constructed to date are based on 
on the $E_8\times E_8$ heterotic--string compactified to four dimensions, 
and the breaking of the $E_8$ symmetry is induced by Wilson lines
\cite{nilles} (for constructions based on the $SO(32)$ heterotic--string
see {\it e.g.} \cite{saul}). 
The $Z_-$ choice in (\ref{changec}) also results in the breaking of the 
right--moving $N=2$ world--sheet symmetry, and is relevant in 
the spinor--vector duality, observed in ref. \cite{fkr}
in the framework of the free fermionic models. Thus, it would be beneficial
to learn how to implement this projection in the orbifold models, which 
will open the way to construct new classes of quasi--realistic orbifold
models, as well as to obtain a realisation of the spinor--vector duality in 
such models.  

\section{Partition functions of NAHE--based models}\label{chalukot}

The partition functions corresponding to the $Z^{4d}_{-}$ and $Z^{4d}_{+}$ vacua 
 after the compactification of the ten dimensional $E_8 \times E_8$
heterotic superstring on the  $SO(12)$ 
lattice at the special point of the moduli space
are given  by 
\beqn
{Z}^{4d}_- =  \frac{({V}_8-{S}_8)}
{{\tau_2}
{(  \eta \overline \eta)}^8} 
&\times&\left[~\left( |O_{12}|^2~+~|V_{12}|^2 ~\right) 
\left( \overline O_{16} \overline  O_{16}+  
\overline C_{16}  \overline C_{16}\right)\right.
\cr
&& + \left( |S_{12}|^2~~+|C_{12}|^2 ~\right) 
\left(  \overline S_{16}  \overline S_{16}+
 \overline V_{16}  \overline V_{16}\right)
\cr
&& + \left( O_{12} \overline  V_{12} + V_{12} \overline  O_{12} \right)
\left(  \overline S_{16}  \overline V_{16} +  
\overline V_{16}  \overline S_{16}\right)
\cr
&& + \left. \left( S_{12}  \overline C_{12} +C_{12}  \overline S_{12} \right)
\left(  \overline O_{16}  \overline C_{16} +  
\overline C_{16}  \overline O_{16} \right) \right] \,,
\label{zminus4d}
\eeqn
and
\beq
{Z}^{4d}_+=\frac{({V}_8-{S}_8)}{{\tau_2}
{(  \eta \overline \eta)}^8} \left[|O_{12}|^2+|V_{12}|^2+
|S_{12}|^2+|C_{12}|^2\right]
\left( \overline  O_{16} + \overline S_{16}\right) 
\left(  \overline O_{16} +  \overline S_{16}
\right) \,, \label{zplus4d}
\eeq
depending on the sign of the discrete torsion.
Here we have written ${Z}_{\pm}$ in terms of level-one
${\rm SO} (2n)$ characters (see, for instance \cite{Angelantonj:2002ct})
\beqn
O_{2n} &=& {\textstyle{1\over 2}} \left( {\vartheta_3^n \over \eta^n} +
{\vartheta_4^n \over \eta^n}\right) \,,
\nonumber \\
V_{2n} &=& {\textstyle{1\over 2}} \left( {\vartheta_3^n \over \eta^n} -
{\vartheta_4^n \over \eta^n}\right) \,,
\nonumber \\
S_{2n} &=& {\textstyle{1\over 2}} \left( {\vartheta_2^n \over \eta^n} +
i^{-n} {\vartheta_1^n \over \eta^n} \right) \,,
\nonumber \\
C_{2n} &=& {\textstyle{1\over 2}} \left( {\vartheta_2^n \over \eta^n} -
i^{-n} {\vartheta_1^n \over \eta^n} \right) \,
\label{thetacharacters}
\eeqn
and the expressions of the form ${|O_{12}|}^2$ mean $O_{12}\overline O_{12}$. 
These two models can actually be connected by the orbifold \cite{partitions}
\beq
{Z}_- = {Z}_+ / a \otimes b \,,
\label{zminusfromzplus}
\eeq
with
\beqn 
a &=& (-1)^{F_{\rm L}^{\rm int} + F_\xi^1} \,,
\nonumber \\
b &=& (-1)^{F_{\rm L}^{\rm int} + F_\xi^2} \,. \label{orbzpm}
\eeqn
Here $F_{\rm L}$ is the fermion number for the ``left''
component in the expression 
of the internal lattice i.e., the only nontrivial action of this operator
is $F_{\rm L} S_{12}= S_{12}$ and $F_{\rm L} C_{12}= C_{12}$. 
The operators
$F_\xi^1$ and $ F_\xi^2$ are fermion number operators in the
first and the second gauge factors respectively.
The orbifold projection given in eqs (\ref{zminusfromzplus}) and (\ref{orbzpm})
is defined at the free fermionic point in the moduli space
since $Z^{4d}_+$ and $Z^{4d}_-$ are expressed at this point. However, it can be
generalised to an arbitrary point in the moduli space and hence can be
employed to construct orbifold models that originate from the $Z^{4d}_-$ 
partition function, in analogy to the case in the free fermionic
constructions. 

The analogous partition function
$Z^{9d}_+$ in the case of one compactified dimension at a general 
point of moduli space is 
given by 
\beq 
{ Z}_+^{9d} =  \frac{({V}_8-{S}_8)}{{\tau_2}^\frac{7}{2}{(  \eta \overline \eta)}^8}    \,  \Lambda_{m,n} \,
		\left( \overline  O_{16} +  \overline S_{16} \right)
		\left(  \overline O_{16} +  \overline S_{16} \right)\,,
\label{zplusin9d}
\eeq
where 
\beq
\Lambda_{m,n} = q^{{\alpha ' \over 4}p_L^2} 
 \bar q^{{\alpha ' \over 4} p_{\rm R}^2}, \quad p_{\rm L,R} = {m \over R} \pm {n R \over \alpha '}
\eeq
applying the orbifold projections 
\beqn
Z_2~:~g &=& (-1)^{F_{\xi^1}}\delta_1 \,,
\nonumber \\
Z_2^\prime~:~g^{\prime} &=& (-1)^{F_{\xi^2}}\delta_1 \,. \label{deltaorbifold}
\eeqn
where 
\begin{equation} \label{shift1}
\delta_1 ~:~ X^9 \rightarrow X^9 +\pi R^9~, \quad \Rightarrow ~~
\Lambda_{mn} ~\rightarrow~ (-1)^m\Lambda_{mn}~.
\end{equation}
 in $Z_+^{9d}$ produces the $Z_-^{9d}$ partition function given by
\beqn 
{ Z}_-^{9d} =  \frac{({V}_8-{S}_8)}{{\tau_2}^\frac{7}{2}
{(  \eta \overline \eta)}^8}~\left[ \right. 
& & \left. \Lambda_{2m,n} \,~~~~~~\left(\overline O_{16}\overline O_{16}+
\overline C_{16}\overline C_{16}\right) \right. \nonumber\\
 &+& \left. \Lambda_{2m+1,n} \, ~~~\left(\overline S_{16}\overline S_{16}~+
\overline V_{16}\overline V_{16}\right)\right. \nonumber\\
 &+& \left. \Lambda_{2m,n+{1\over2}} \, ~~~\left(\overline S_{16}\overline 
V_{16}+\overline V_{16}\overline S_{16}\right)\right. \nonumber\\
 &+&  \left. \Lambda_{2m+1,n+{1\over2}} \, \left(\overline O_{16}\overline 
C_{16}+\overline C_{16}\overline O_{16}\right)\right]~.\nonumber\\
\label{zminus9d}
\eeqn
Let us note that the internal fermion number
operator which appears in (\ref{orbzpm})
is now replaced by the shift along the compact dimension, given in
(\ref{shift1}).

We note here that the shift given by delta differs from the shifts that were 
found in ref. \cite{partitions} to reproduce the partition function of the
$SO(12)$ lattice at the maximally symmetric 
free fermionic point, given in eq.(\ref{zplus4d}).
Since the string contains momentum and winding modes along a
compactified coordinate, one can shift
 the coordinate along either, and also allow shifts that mix the momentum
and winding modes \cite{vwaaf}.
Indeed, the precise identification of the lattice at the 
free fermionic point is obtained for shifts that mix momentum and winding
modes. Here we restrict ourselves to the simpler shifts and incorporation
of other shifts is left for a future work.

Let us consider  partition functions (\ref{zplusin9d}) and
(\ref{zminus9d}) at the special (free fermionic)
point of the moduli space i.e.,   
\beq
Z^{9d}_+=\frac{({V}_8-{S}_8)}{{\tau_2}^\frac{7}{2}{(\eta\overline\eta)}^8}
\left(|O_2|^2+|V_2|^2+|S_2|^2+|C_2|^2\right)
\left( \overline O_{16} + \overline S_{16}\right) \left( \overline O_{16} +
\overline S_{16}\right) 
\label{zplus9d}
\eeq
and
\ba
Z^{9d}_-=\frac{({V}_8-{S}_8)}{{\tau_2}^\frac{7}{2}
{(  \eta \overline \eta)}^8}  &\times& \left[\left(|O_2|^2+|V_2|^2\right)
\left(\overline O_{16} \overline O_{16}+ \overline C_{16}
\overline C_{16}\right) \right. \cr\cr
&& + 
\left(|S_2|^2+|C_2|^2\right) \left( \overline S_{16} \overline S_{16}+
\overline V_{16} \overline V_{16}\right) 
\cr
\cr
&& + \left(O_2\overline{V}_2+V_2\overline{O}_2\right)
\left( \overline S_{16} \overline V_{16} + \overline V_{16}
\overline S_{16}\right) \cr\cr
&&  + \left. \left(S_2\overline{C}_2+C_2\overline{S}_2\right)
\left( \overline O_{16} \overline C_{16} + \overline C_{16}
\overline O_{16} \right) \right] .
\label{zminus9dchara}
\ea
In this case the $Z_2\times Z_2^\prime$ orbifold operation is again given by
 (\ref{orbzpm}) where $F_L^{\rm int}$ 
acts on $S_2$ and $C_2$ of the internal one -- dimensional lattice. 
Equations (\ref{zplus9d})
and (\ref{zminus9dchara}) therefore have the same structure as the corresponding
four dimensional equations (\ref{zminus4d}) and (\ref{zplus4d}) respectively. 

We now comment on the ten dimensional case. In terms of the free fermionic
formalism both $Z_+$ and $Z_-$ models are generated by the set of basis vectors 
$\{1,\xi_1,\xi_2\}$. The $S$--sector is obtained as a combination of the
basis vectors with $S=1+\xi_1+\xi_2$. As a result the 
GGSO coefficient $c{S \choose \xi_1}$ is proportional to $c{\xi_2 \choose \xi_1}$.
Consequently, the choice of the GGSO coefficient 
$$c{\xi_2 \choose \xi_1}=-1~,$$
reduces the $E_8\times E_8$ gauge symmetry to $SO(16)\times SO(16)$
as well as 
projects out the space--time supersymmetry generator from the $S$--sector.
In terms of $c{\xi_1 \choose \xi_2}$ the free fermionic partition function takes
the form
\ba \label{pf10d}
Z^{10d}&=& \frac{1}{{\tau_2}^4{(  \eta \overline \eta)}^8}
{1\over 2^3}\left\{\left( \theta_3^4-\theta_4^4-
\theta_2^4-\theta_1^4\right) \left(\bar{\theta}_3^{16}+\bar{\theta}_4^{16} 
+ \bar{\theta}_2^{16} + \bar{\theta}_1^{16}\right)\right.
\cr
\cr
&&+\left[\theta_3^4-\theta_4^4-c{\xi_1\choose \xi_2}
\left(\theta_2^4+\theta_1^4\right)\right]
\left[\bar{\theta}_3^8\bar{\theta}_4^8+
\bar{\theta}_4^8\bar{\theta}_3^8+
c{\xi_1\choose \xi_2}\left(\bar{\theta}_2^8\bar{\theta}_1^8+
\bar{\theta}_1^8\bar{\theta}_2^8\right)\right]
\cr
\cr
&&+\left[\theta_3^4-c{\xi_1\choose \xi_2}\left(\theta_4^4+
\theta_1^4\right)-\theta_2^4\right]\left[\bar{\theta}_2^8\bar{\theta}_3^8+
\bar{\theta}_3^8\bar{\theta}_2^8+c{\xi_1\choose \xi_2}
\left(\bar{\theta}_4^8\bar{\theta}_1^8+
\bar{\theta}_1^8\bar{\theta}_4^8\right)\right]
\cr
\cr
&&\left.\left[c{\xi_1\choose \xi_2}\left(\theta_3^4-\theta_1^4\right)-
\theta_4^4-\theta_2^4\right]\left[\bar{\theta}_2^8\bar{\theta}_4^8+
\bar{\theta}_4^8\bar{\theta}_2^8+c{\xi_1\choose \xi_2}
\left(\bar{\theta}_3^8\bar{\theta}_1^8+
\bar{\theta}_1^8\bar{\theta}_3^8\right)\right]\right\}\nonumber\\
\ea
The choice $c{\xi_1\choose \xi_2}=1$ of the GGSO coefficient corresponds to
the partition function $Z_+$ and the choice 
$c{\xi_1\choose \xi_2}=-1$ corresponds
to $Z_-$.
In terms of the characters of eq. (\ref{thetacharacters}) $Z_+$ takes
the form
\beq
 Z_+^{10d}= \frac{({V}_8-{S}_8)}{{\tau_2}^4{(  \eta \overline \eta)}^8}
\left( \overline O_{16} + \overline S_{16}\right)
\left( \overline O_{16} + \overline S_{16}
\right) 
\label{z10dplus}
\eeq
and $Z_-^{10d}$
\ba
 Z^{10d}_-~=~  \frac{1}{{\tau_2}^4{(  \eta \overline \eta)}^8}~~
&\left[ \right. & 
V_8 \left(\overline O_{16} \overline O_{16}+ 
           \overline C_{16} \overline C_{16}\right)
\cr
- && 
S_8 \left( \overline S_{16} \overline S_{16}+
            \overline V_{16} \overline V_{16}\right)
\cr
+ &&  
O_8 \left( \overline S_{16}  \overline V_{16} + 
           \overline V_{16}  \overline S_{16}\right) 
\cr
- && \left.
C_8 \left( \overline O_{16} \overline C_{16}  + 
           \overline C_{16} \overline O_{16} \right) \right]. 
\label{z10dminus}
\ea
The projection to be implemented in $Z_+$  to obtain $Z_-$ is 
\beq 
\frac{1+(-1)^{F+F_{\xi_1}}}{2}\times \frac{1+(-1)^{F+F_{\xi_2}}}{2},
\label{tendprojection}
\eeq
with $F$ being the space-time fermion number and 
$F_{\xi_{1,2}}$ as before.  
Therefore the  projection in the ten dimensional case
reproduces the partition function
of the non--supersymmetric $SO(16)\times SO(16)$ heterotic string.

In order to see what happens for the case of compactification to nine dimensions
one can rewrite  the partition functions 
(\ref{zplus9d}) and (\ref{zminus9dchara})
in the form similar to (\ref{pf10d}) i.e.,
therms of $\theta$ functions. The one can show
that the difference between these partition functions 
is due to a choice of the discrete torsion $c{\xi_1\choose \xi_2}= \pm 1$.
One can conclude that in the free fermionic construction
the same phase that reduces
the gauge symmetry in the compactified model from  $E_8\times E_8$ 
to $SO(16)\times SO(16)$, uniquely in the uncompactified case also projects out 
the space--time supersymmetry generator. The natural question is therefore
what is the relation in the compactified case between the supersymmetric an
non--supersymmetric models. We turn to address this question in the next 
section. 

\section{Interpolations}\label{interpolazia}

To explore the connection between the two models we explore the following
compactifications. First we study the compactification of the heterotic
$E_8\times E_8$ on a circle $S_1$ moded by $Z_2\times Z_2^\prime$
orbifold as in  (\ref{shift1}).
 The resulting theory is a heterotic--string 
with $N=1$ supersymmetry and $SO(16)\times SO(16)$ gauge symmetry.
Schematically, we have
\beq
{{\left(\hbox{Het } 
E_8\times E_8 ~\hbox{on~} S_1 \right)}\over{Z_2\times Z_2^\prime}}=
\hbox{het } SO(16)\times SO(16) ~\hbox{on~} S_1 ~\hbox{with } N=1 
\quad \hbox{SUSY}
\eeq 
Next, we compactify this heterotic $SO(16)\times SO(16)$ model on 
another circle $S_1^\prime$ moded
by $Z_2^{\prime}$ with the shift  
\beq
Z_2^{\prime \prime}~:~g^{\prime \prime}=(-1)^{F+F_{\xi_1}+F_{\xi_2}}\delta_2
\label{z2primeprime}
\eeq
and 
\beq
\delta_2 ~:~ X_8 ~\rightarrow~ X_8+\pi R_8~~\Rightarrow ~~
\Gamma_{mn} ~\rightarrow~ (-1)^m\Gamma_{mn}~.
\label{delta2}
\eeq
The fermion numbers $F_{\xi_1}$ and $F_{\xi_2}$ in eq. (\ref{z2primeprime})
are as before, whereas the fermion number $F$ refers to the 
space--time fermion number.
We then demonstrate that the limit $R_8 \rightarrow0$ of the
$SO(16)\times SO(16)$ non--supersymmetric heterotic string in 
eight dimensions produces the partition function of the 
ten dimensional $SO(16)\times SO(16)$ non--supersymmetric
heterotic string compactified on a circle $S_1$, moded by the
$Z_2\times Z_2^\prime$ orbifold. Alternatively, 
the decompactification limit $R_8 \rightarrow\infty$ of the non--supersymmetric
heterotic string compactified to eight dimensions, interpolates to 
the a supersymmetric $SO(16)\times SO(16)$ heterotic--string in nine
dimensions, moded by the $Z_2\times Z_2^\prime$ orbifold.

To demonstrate these interpolations 
in a manner similar to \cite{itoyamataylor,Blum:1997gw} 
we start with the partition function of 
the $E_8\times E_8$ heterotic--string compactified on one $S_1$. The partition
function of this vacuum is given in eq. (\ref{zplus9d}). We then implement the 
$Z_2\times Z_2^\prime$ orbifold projection given in eq. (\ref{orbzpm}). 
Operating with the $Z_2\times Z_2^\prime$ introduces a discrete 
torsion, $\epsilon=\pm 1$, in the partition function, which is an undefined sign
between independent modular orbits. The  choice of the negative 
sign produces a supersymmetric nine dimensional string vacuum with $SO(32)$
gauge group, with 
\beq
\overline Z^{9d}_-= \frac{(V_8-S_8)}{{\tau_2}^\frac{7}{2}{(  \eta \overline \eta)}^8} 
\left[
({\overline O}_{32}P_+ + {\overline S}_{32}P_-)\Lambda+
({\overline V}_{32}P_- + {\overline C}_{32}P_+)\Lambda_{1\over2}\right]
\eeq
where  we introduced the notation 
\beq
P_\pm={1\over2}(1\pm(-1)^m)
\label{ppm}
\eeq
and
\beqn
\Lambda          ~~ & \equiv & \Lambda_{m,n}\nonumber\\
\Lambda_{1\over2} & \equiv & \Lambda_{m,n+{1\over2}}
\label{lamlam1over2}
\eeqn
We note that the $\pm$ index here refers to the choice of the 
discrete torsion, $\epsilon=\pm1$,
and differs from the $\pm$ index used previously
which referred to the choice of the sign in eq. (\ref{changec}).
The choice $\epsilon=+1$ yields
\beqn
\overline Z^{9d}_+&=& \frac{(V_8-S_8)}{{\tau_2}^\frac{7}{2}{(  \eta \overline \eta)}^8}
\left\{ \right. 
 \left[
  ({\overline O}_{16}{\overline O}_{16}+
   {\overline C}_{16}{\overline C}_{16})P_+ + 
  ({\overline S}_{16}{\overline S}_{16}+
   {\overline V}_{16}{\overline V}_{16})P_- 
 \right]\Lambda
+ \nonumber\\
 &&  \left. ~~~\left[
  ({\overline C}_{16}{\overline O}_{16}+
   {\overline O}_{16}{\overline C}_{16})P_+ + 
  ({\overline V}_{16}{\overline S}_{16}+
   {\overline S}_{16}{\overline V}_{16})P_- ~
 \right]\Lambda_{1\over2}~~~
\right\}
\label{zdtplus}
\eeqn
which is a generalization of (\ref{zminus9dchara}) to an arbitrary point in the moduli space.
The partition function in eq. (\ref{zdtplus}) is our starting point for
interpolating between the relevant string vacua discussed above. We next
mod out the partition function given in eq. (\ref{zdtplus}) by the 
$Z_2^{\prime\prime}$ orbifold given in eq. (\ref{z2primeprime}).
The resulting partition function is given by
\beqn
\overline Z^{8d}_+=\frac{1}{{\tau_2}^3{(  \eta \overline \eta)}^8} 
[\left\{ \right. & 
 V_8 &
  \left[
   ({\overline O}_{16}{\overline O}_{16}+
    {\overline C}_{16}{\overline C}_{16})P_+ + 
   ({\overline S}_{16}{\overline S}_{16}+
    {\overline V}_{16}{\overline V}_{16})P_- ~
  \right]\Lambda
\\ \nonumber
- & S_8  &  \left. 
  \left[
  ({\overline C}_{16}{\overline O}_{16}+
   {\overline O}_{16}{\overline C}_{16})P_+ + 
  ({\overline V}_{16}{\overline S}_{16}+
   {\overline S}_{16}{\overline V}_{16})P_- ~
 \right]\Lambda_{1\over2}~
\right\}P_+\Gamma \\ \nonumber
+ 
\left\{ \right. & 
 V_8 &
  \left[
   ({\overline O}_{16}{\overline C}_{16}+
    {\overline C}_{16}{\overline O}_{16})P_+ + 
   ({\overline V}_{16}{\overline S}_{16}+
    {\overline S}_{16}{\overline V}_{16})P_- ~
  \right]\Lambda_{1\over2}~
\\ \nonumber
- & S_8  &  \left. 
  \left[
  ({\overline O}_{16}{\overline O}_{16}+
   {\overline C}_{16}{\overline C}_{16})P_+ + 
  ({\overline S}_{16}{\overline S}_{16}+
   {\overline V}_{16}{\overline V}_{16})P_- ~
 \right]\Lambda~~~
\right\}P_-\Gamma \\ \nonumber
+ 
\left\{ \right. & 
 O_8 &
  \left[
   ({\overline O}_{16}{\overline C}_{16}+
    {\overline C}_{16}{\overline O}_{16})P_- + 
   ({\overline V}_{16}{\overline S}_{16}+
    {\overline S}_{16}{\overline V}_{16})P_+ ~
  \right]\Lambda_{1\over2}~
\\ \nonumber
- & C_8  &  \left. 
  \left[
  ({\overline O}_{16}{\overline O}_{16}+
   {\overline C}_{16}{\overline C}_{16})P_- + 
  ({\overline V}_{16}{\overline V}_{16}+
   {\overline S}_{16}{\overline S}_{16})P_+ ~
 \right]\Lambda~~~
\right\}P_+\Gamma_{1\over2}~ \\ \nonumber
+ 
\left\{ \right. & 
 O_8 &
  \left[
   ({\overline O}_{16}{\overline O}_{16}+
    {\overline C}_{16}{\overline C}_{16})P_- + 
   ({\overline V}_{16}{\overline V}_{16}+
    {\overline S}_{16}{\overline S}_{16})P_+ ~
  \right]\Lambda~~~
\\ \nonumber
- & C_8  &  \left. 
  \left[
  ({\overline O}_{16}{\overline C}_{16}+
   {\overline C}_{16}{\overline O}_{16})P_- + 
  ({\overline V}_{16}{\overline S}_{16}+
   {\overline S}_{16}{\overline V}_{16})P_- ~
 \right]\Lambda_{1\over2}~
\right\}P_+\Gamma_{1\over2}]~  \\ \nonumber
& \label{zdtplusnonsusy}
\eeqn
We can now study the various limits as $R_8 \rightarrow0$ and 
$R_8 \rightarrow\infty$, where $R_8$ is the radius of the 
$\Gamma$ lattice which accounts for the spontaneous supersymmetry
breaking. These limits are: 

\beq
\begin{tabular}{c|c}
 ~~~ $R_8 \rightarrow\infty$ & ~~ $ R_8 \rightarrow 0$ \\
\hline
\hline
 ~~~ ${\Gamma}\rightarrow 1$                
& ~~~~${\Gamma}\rightarrow 1$  \\
  $(-1)^m{\Gamma}\rightarrow 0$   ~~~       
& ~$(-1)^m{\Gamma}\rightarrow 1$ ~~~  \\
 ~~ ${\Gamma}_{1\over2}\rightarrow 0$        
& ~~~${\Gamma}_{1\over2}\rightarrow 1$  \\
  $(-1)^m{\Gamma}_{1\over2}\rightarrow 0$ ~~~~~ 
& $(-1)^m{\Gamma}_{1\over2}\rightarrow 1$~~~~  \\
\hline
\end{tabular}
\eeq
and $ \frac{1}{{\tau_2}^3{(  \eta \overline \eta)}^8} 
\rightarrow \frac{1}{{\tau_2}^\frac{7}{2}{(  \eta \overline \eta)}^8}$
 
Therefore, in the limit $R_8 \rightarrow\infty$ 
we recover the partition function in 
eq. (\ref{zdtplus}), whereas the limit $R_8 \rightarrow 0$ yields 
\beqn
\overline Z_{R_8 \rightarrow0}^{9d}= \frac{1}{{\tau_2}^\frac{7}{2}{(  \eta \overline \eta)}^8} 
( \left\{ \right. & 
 V_8 &
  \left[
   ({\overline O}_{16}{\overline O}_{16}+
    {\overline C}_{16}{\overline C}_{16})P_+ + 
   ({\overline S}_{16}{\overline S}_{16}+
    {\overline V}_{16}{\overline V}_{16})P_- ~
  \right]\Lambda
\\ \nonumber
- & S_8  &  \left. 
  \left[
  ({\overline C}_{16}{\overline O}_{16}+
   {\overline O}_{16}{\overline C}_{16})P_+ + 
  ({\overline V}_{16}{\overline S}_{16}+
   {\overline S}_{16}{\overline V}_{16})P_- ~
 \right]\Lambda_{1\over2}~\right\} \\ \nonumber
+ 
\left\{ \right. & 
 O_8 &
  \left[
   ({\overline O}_{16}{\overline C}_{16}+
    {\overline C}_{16}{\overline O}_{16})P_- + 
   ({\overline V}_{16}{\overline S}_{16}+
    {\overline S}_{16}{\overline V}_{16})P_+ ~
  \right]\Lambda_{1\over2}~
\\ \nonumber
- & C_8  &  \left. 
  \left[
  ({\overline O}_{16}{\overline O}_{16}+
   {\overline C}_{16}{\overline C}_{16})P_- + 
  ({\overline V}_{16}{\overline V}_{16}+
   {\overline S}_{16}{\overline S}_{16})P_+ ~
 \right]\Lambda~~~
\right\})~ \\ \nonumber
& \label{zdtplusnonsusyrhotozero}
\eeqn
We need to understand now what theory this partition corresponds to 
and how it is connected to the non--supersymmetric non--tachyonic 
$SO(16)\times SO(16)$, 
or directly to the supersymmetric $E_8\times E_8$ heterotic--string theory. 
For this we consider the partition function of the ten dimensional 
$E_8\times E_8$ heterotic--string given in eq. (\ref{z10dplus})
moded by the orbifold given in eq. (\ref{tendprojection}).
The resulting partition function is given in eq. (\ref{z10dminus})
and is nonsupersymmetric $SO(16) \times SO(16)$ heterotic string.
After that  we compactify the non--supersymmetric 
$SO(16)\times SO(16)$ heterotic string on a circle $S_1$ and project with the
freely acting $Z_2\times Z_2^\prime$ orbifold given in eq. 
(\ref{deltaorbifold}).
The partition function in this case,with the  discrete torsion 
$\epsilon=-1$ exactly reproduces  the one given in eq. 
(\ref{zdtplusnonsusyrhotozero}), 
which is obtained from eq. (\ref{zdtplusnonsusy}) in the limit
$ R_8 \rightarrow0$. 

Alternatively to the above derivation we can interpolate the vacua at the 
special point in the moduli space, with the lattice given in terms of 
the characters 
of eq. (\ref{thetacharacters}). Starting, say, from the $E_8\times E_8$
heterotic--string compactified to six dimensions, with
\beq
 {Z}^{6d}_+= \frac{( V_8- S_8)}{{\tau_2}^2{(  \eta \overline \eta)}^8} 
\left[|O_{8}|^2+|V_{8}|^2+|S_{8}|^2+|C_{8}|^2\right]
\left( \overline  O_{16} + \overline S_{16}\right) \left(  \overline O_{16} + 
\overline S_{16}\right) \,, \label{zplus6d}
\eeq
Further we perform a compactification on a circle $S_1$ with a $Z_2 \times Z_2^\prime$
obifold action defined in (\ref{zminusfromzplus}), with $F_L S_8 = S_8$ and $F_L C_8 = C_8$
since now we  are dealing with a four dimensional internal lattice. Further we compactify the
model on another circle with  $Z_2^{\prime \prime}$ orbifold action defined in (\ref{z2primeprime}).
As it could be expected the result of this procedure is the same as
in the case  if the compactification to eight dimensions which was discussed above in this chapter with the following substitutions
\beqn
P_+ \Lambda~~ &~\rightarrow~ & \vert O_8\vert^2 + \vert V_8\vert^2\nonumber\\
P_- \Lambda~~ &~\rightarrow~ & \vert S_8\vert^2 + \vert C_8\vert^2\nonumber\\
P_- \Lambda_{1\over2}~ &~\rightarrow~ &  O_8{\bar V_8} + V_8{\bar O_8}\nonumber\\
P_+ \Lambda_{1\over2}~ &~\rightarrow~ &  S_8{\bar C_8} + C_8{\bar S_8}.
\label{substitutions8lattice}
\eeqn

To summarise, we note that the choice of the discrete phase  in the free fermionic
models is realised as an orbifold in the bosonic formalism. It is represented
as a fermion number acting in the gauge sector, coupled with a shift in 
the internal lattice or, alternatively, with a fermion number
acting in the internal lattice at the enhanced symmetry point. 
The choice of the specific free phase in $c{\xi_1\choose \xi_2}=\pm1$
determines many of the phenomenological properties of the free fermion
models. Understanding how the projection induced by this
phase is realised in the bosonic language will facilitate
the reproduction of these features in orbifold models.
In the ten dimensional free fermion model the choice
$c{\xi_1\choose \xi_2}=-1$ projects out $N=1$ supersymmetry.
On the other hand, in lower dimensional free fermion models with 
$c{\xi_1\choose \xi_2}=-1$ $N=1$ space--time supersymmetry
is preserved, and the connection between the models is
obscured. As we demonstrated here the bosonic
representation of the models clarifies the situation and
elucidated how the different cases are connected by interpolations.
This demonstrates how the fermionic and bosonic representations of the
string vacua provide complementary tools to reveal their
properties.

\section{Conclusions}\label{maskanot}

String theory provides a unique framework to study how a spin 2 quantum 
field theory
may be consistently incorporated with lower spin quantum field theories. 
As such it provides a unique arena to explore how the phenomenological 
properties
of the Standard Model may arise from a theory of quantum gravity, 
and is hence
the leading contender for a unified theory of gravity and the gauge 
interactions.
Indeed string theory gives rise to a multitude of vacua that can in principle 
be relevant for low scale phenomenology. Among those the models constructed 
in the free fermionic formulations admit particularly appealing structure. 
The existence of a plethora of vacua would seem problematic from the low 
energy effective quantum field theory point of view. However, the lesson from
the existence of symmetries under various duality transformations is 
that vacua 
that appear to be disconnected in the effective field theory limit may in fact be
connected in the string theory due to the existence of the massive spectrum.

A key ingredient in the construction of the phenomenological free
fermionic models is the GGSO phase, eq. (\ref{changec}). In the 
quasi--realistic four dimensional models this choice of phase
accounts for the reduction of the GUT symmetry from $E_6~\rightarrow SO(10)$,
which results in Higgs--matter splitting \cite{higgsmattersplit}. Uniquely, 
in the ten dimensional case this choice of phase also projects out the 
space--time supersymmetry generator. While the free fermionic models possess
phenomenologically attractive properties, it is of interest to obtain
a representation of these models in a bosonic, \ie~ orbifold, formulation.
The reason being that the bosonic representation can reveal the geometrical
structures underlying these models and hence to probe possible dynamical 
vacuum selection scenarios. Hence, it is of interest to learn how the 
breaking induced by the phase given in eq. (\ref{changec}) operates in the orbifold models. 
In general, we observe that this breaking is induced in the bosonic formalism
by coupling the fermion number of the observable and hidden gauge degrees of 
freedom with a fermion number acting asymmetrically on the internal lattice at the self--dual
point in the moduli space, or on the
space--time fermions. Alternatively, the action in the internal dimensions
can be represented as a mod two shift in the compactified coordinates. 
In the ten dimensional case the only possible internal action is on the 
space--time fermions, which breaks supersymmetry. 
We further addressed in this paper the question of how the models with
the different actions of the fermion number in the internal spaces
may be connected. This question is similar to the connection between 
the $E_8\times E_8$ and $SO(16)\times SO(16)$ heterotic strings,
obtained by interpolations \cite{itoyamataylor}.
Indeed, we demonstrated in section
\ref{interpolazia} that the models produced by these differing 
actions can be connected by interpolations, \ie~ that the relevant partition 
functions are reproduced in the appropriate limits. 
The phenomenologically appealing properties of the free fermionic
models motivates their exploration in further depth. All the models 
studied to date in this context are supersymmetric and stable.
However, it is plausible that unstable and nonsupersymmetric vacua
can be instrumental for probing questions pertaining to 
dynamical vacuum selection scenarios.
It is therefore vital to explore how the nonsupersymmetric models are
incorporated in this picture,
as well as in that of M--theory \cite{ft,martinec}.
Understanding the bosonic
equivalence of the free fermionic models, and their embedding
in cosmological string solutions, offers the possibility of 
developing such a dynamical string vacuum selection scenario.
We hope to return to these questions in the future. 

\bigskip
\medskip
\leftline{\large\bf Acknowledgements}
\medskip

We would like to thank Carlo Angelantonj for
collaboration at the initial stage of the project and discussions
throughout and
Elisa Manno, Thomas Mohaupt and Cristina Timirgaziu
for useful discussions. 
AEF thanks the Galileo Galilei Institute for hospitality.
This work is supported in part by a STFC rolling grant ST/G00062X/1.



\bigskip
\medskip

\bibliographystyle{unsrt}

\vfill\eject
\end{document}